\documentclass[aps,prx,superscriptaddress,preprint,longbibliography,amsmath]{revtex4-1}

\usepackage{gensymb}
\usepackage{graphicx}

 \usepackage{color}
 \usepackage{hyperref} 
\usepackage{cleveref}

\usepackage{amsmath}
\usepackage{amssymb}
\usepackage{pifont}
\usepackage{latexsym}
\usepackage{dsfont }
\usepackage{float}
\usepackage{multirow}
\usepackage{verbatim}

\usepackage{standalone}
\usepackage{subfiles}

\usepackage{pdfpages}

\makeatletter
\AtBeginDocument{\let\LS@rot\@undefined}
\makeatother

\def\bk{{\bold{k}}}
\def\br{{\bold{r}}}

\def\up{{\uparrow}}
\def\down{{\downarrow}}

\def\m1{{^{-1}}}

\begin{document}

\title{Generalized Anderson's theorem for superconductors \\
derived from topological insulators}

\author{Lionel Andersen}
\thanks{These two authors contributed equally.}
\affiliation{Physics Institute II, University of Cologne, 50937 K\"oln, Germany}

\author{Aline Ramires}
\thanks{These two authors contributed equally.}
\affiliation{Max Planck Institute for the Physics of Complex Systems, Dresden, 01187, Germany}
\affiliation{ICTP-SAIFR, International Centre for Theoretical Physics - South American Institute for Fundamental Research, S\~{a}o Paulo, SP, 01140-070, Brazil}
\affiliation{Instituto de F\'{i}sica Te\'{o}rica - Universidade Estadual Paulista, S\~{a}o Paulo, SP, 01140-070, Brazil}

\author{Zhiwei Wang}
\author{Thomas Lorenz}
\affiliation{Physics Institute II, University of Cologne, 50937 K\"oln, Germany}

\author{Yoichi Ando}
\affiliation{Physics Institute II, University of Cologne, 50937 K\"oln, Germany}

\date{\today}

\begin{abstract}

A well-known result in unconventional superconductivity is the fragility of nodal superconductors against nonmagnetic impurities. Despite this common wisdom, Bi$_2$Se$_3$-based topological superconductors have recently displayed unusual robustness against disorder. Here we provide a theoretical framework which naturally explains what protects Cooper pairs from strong scattering in complex superconductors. Our analysis is based on the concept of \emph{superconducting fitness} and generalizes the famous Anderson's theorem into superconductors having multiple internal degrees of freedom. For concreteness, we report on the extreme example of the Cu$_x$(PbSe)$_5$(Bi$_2$Se$_3$)$_6$ superconductor, where thermal conductivity measurements down to 50 mK not only give unambiguous evidence for the existence of nodes, but also reveal that the energy scale corresponding to the scattering rate is orders of magnitude larger than the superconducting energy gap. This provides a most spectacular case of the generalized Anderson's theorem protecting a nodal superconductor.

\vspace{5mm}

\begin{flushleft}
{\bf One-sentence summary: Cooper pairs in unconventional superconductors having extra 
internal degrees of freedom are protected in an unexpected way.}
\end{flushleft}

\end{abstract}





\maketitle


Unconventional superconductors distinguish themselves from conventional ones by breaking not only $U(1)$ gauge, but also additional symmetries, usually reducing the point group associated with the normal-state electronic fluid \cite{SigristAIP2005}. This extra symmetry reduction stems from the development of order parameters with nontrivial form factors, typically introducing point or line nodes in the excitation spectra \cite{StewartAIP17}. Nodal gap structures are especially known to give rise to power-law behavior in transport and thermodynamic quantities, which can be clearly detected in experiments, and are established as a key signature of unconventional superconductivity \cite{Hussey2002, Shakeripour2009, Lee1993, Graf1996, Taillefer1997, Proust2002, Izawa2001, Nagai2008}. However, nodal structures are also known to make superconductivity fragile in the presence of impurities \cite{BalianPR1963}, and many unconventional superconductors  have actually been shown to be extremely sensitive to disorder \cite{BalianPR1963, MackenzieRMP2003,MackenziePRL1998, DalichaouchPRL1995}. 

Against all odds, the superconductivity in Bi$_2$Se$_3$-based materials was recently reported to present unusual robustness against disorder \cite{Kriener2012, Smylie2017}, despite showing nematic properties which point to unconventional topological superconductivity \cite{Sato2017}. Here we report a striking observation that the Cu$_x$(PbSe)$_5$(Bi$_2$Se$_3$)$_6$ (CPSBS) superconductor \cite{Sasaki2014}, which also show nematic properties \cite{Andersen2018}, gives unambiguous evidence for the existence of gap nodes while the scattering rate is more than an order of magnitude larger than the gap, a circumstance where nodal superconductivity is completely suppressed according the common wisdom. To understand this apparent puzzle, we generalize Anderson's theorem \cite{AndersonJPCS1959, SuhlPR1959} to complex superconducting (SC) materials encoding extra internal degrees of freedom (DOF), such as orbitals, sublattices, or valleys. It turns out that as long as the pairing interaction is isotropic, superconductors having a momentum-dependent gap structure, which manifests itself in the band basis, are generically protected from nonmagnetic scattering that do not mix the internal DOF. Our analysis is based on the concept of \emph{superconducting fitness}, a useful tool for understanding the robustness of SC states involving multiple DOF.

\begin{flushleft}
{\bf Generalizing Anderson's Theorem}
\end{flushleft}

We start from generalizing the Anderson's theorem to superconductors having extra internal DOF. To address the effects of impurities in such superconductors, it is useful to consider a Bogoliubov-de Gennes Hamiltonian
\begin{eqnarray}\label{Eq:HBdG}
H_{BdG} (\bk)&=& \Psi_\bk^\dagger \begin{pmatrix}
 H_{0}(\bk)& \Delta(\bk)\\
\Delta^\dagger(\bk) & - H^*_{0} (-\bk)
\end{pmatrix}
\Psi_\bk,
\end{eqnarray}
written in terms of a \emph{multi-DOF Nambu spinor} $\Psi_\bk^\dagger = (\Phi^\dagger_{\bk}, \Phi_{-\bk}^T)$, encoding several DOF within $\Phi^\dagger_{\bk}= (c_{1\bk \up}^\dagger, c_{1\bk \down}^\dagger ,..., c_{n\bk \up}^\dagger, c_{n\bk \down}^\dagger)$. Here $c_{m\bk\sigma}^\dagger$  ($c_{m\bk\sigma}$) creates (annihilates) an electron in the internal DOF $m$ with momentum $\bk$ and spin $\sigma=\{\uparrow,\downarrow\}$. $H_0(\bk)$ is the normal-state Hamiltonian in this multi-DOF basis, which can be parametrized as
\begin{eqnarray}\label{Eq:H0}
H_0(\bk)= \sum_{a,b}  \Phi^\dagger_\bk [h_{ab}(\bk) \tau_a\otimes\sigma_b] \Phi_\bk,
\end{eqnarray}
where $h_{ab}(\bk)$ are momentum-dependent real functions with subscripts $a$ and $b$ corresponding to the extra internal DOF and the spin DOF, respectively. If we focus on the case of two orbitals as the extra internal DOF (as in the Bi$_2$Se$_3$-based superconductors), $\tau_{i}$ and $\sigma_{i}$ ($i=\{1,2,3\}$) are the Pauli matrices to encode the orbital and spin DOFs, respectively, and $\tau_0$ and $\sigma_0$ are identity matrices. In this case there are in principle sixteen parameters $h_{ab}(\bk)$. However, in the presence of time-reversal and inversion symmetries, the number of allowed $h_{ab}(\bk)$ terms is reduced to only five plus $h_{00}(\bk)$, with the associated matrices $ \tau_a\otimes\sigma_b$ forming a set of totally anticommuting matrices (see Table S1 in the Supplementary Materials). 

In Eq. \ref{Eq:HBdG}, $\Delta(\bk)$ is the gap matrix, which can be parametrized in a similar form:
\begin{eqnarray}\label{Eq:Delta}
\Delta(\bk) = \sum_{a,b}  \Phi^\dagger_\bk [d_{ab} (\bk) \tau_a \otimes \sigma_b (i\sigma_2) ]\Phi^*_{-\bk}.
\end{eqnarray}
Here $d_{ab} (\bk)$ denote form factors, which, in general, can have a $\bk$-dependence determined by the pairing mechanism. However, when superconductivity is driven by phonons or by local interactions, the pairing force is isotropic and $d_{ab}$ becomes independent of $\bk$. In the following, having Bi$_2$Se$_3$-based superconductors in mind, we focus on $\bk$-independent $d_{ab}$, because the pairing force is considered to be isotropic in those materials \cite{Fu2010}.

The effects of impurities in multi-DOF superconductors can be understood by calculations similar in spirit as the standard calculations for simple metals \cite{Maki1969,AG2012} (details in the Supplementary Materials), from which we can infer the behavior of the critical temperature, $T_c$, as a function of the effective scattering rate in the normal state, $\hbar \Gamma_{\rm Eff}$. The calculations yield a familiar result, which is now generalized to encode the complexity of the normal and SC states in the multi-DOF basis,
\begin{eqnarray}\label{Eq:Tc}
\log\left(\frac{T_c}{T_c^0}\right) = \Psi\left(\frac{1}{2}\right) - \Psi\left(\frac{1}{2}+\frac{\hbar \Gamma_{\rm Eff}}{2 \pi k_B T_c}\right),
\end{eqnarray}
where $T_c^0$ is the critical temperature of the clean system, $\Psi(x)$ is the digamma function, and $\hbar\Gamma_{\rm Eff}$ encodes all pair-breaking mechanisms through
\begin{eqnarray}
\hbar \Gamma_{\rm Eff}= \frac{1}{4} \left\langle \overline{Tr[\tilde{F}_{C}^\dagger (\Omega_\bk)\tilde{F}_{C}(\Omega_\bk)]} \right\rangle_\bk,
\end{eqnarray}
which is determined solely by the \emph{superconducting fitness} function:
\begin{eqnarray}
F_C(\bk- \bk') = V(\bk-\bk' ) \Delta  - \Delta  V^*(\bk-\bk').
\end{eqnarray}
This expression is valid for $\bk$-independent $\Delta$ matrices of arbitrary dimension. Here $V(\bk-\bk')$ is the matrix impurity scattering potential encoding all DOFs, $\tilde{F}_C(\Omega_\bk) = F_C(\Omega_\bk)/\Delta_0$, $\Delta_0$ is the magnitude of the gap, $\Omega_\bk$ is the solid angle at the Fermi surface, the horizontal bar indicates impurity averaging, and the brackets indicate the average over the Fermi surface. This form of the effective scattering rate accounts for the potentially nontrivial dependences of the pair wavefunctions and scattering processes on the multiple DOFs. It is useful to mention that the superconducting fitness function $F_C(\bk)$ was originally introduced as a measure of the incompatibility of the normal-state electronic structure with the gap matrix, defined as a modified commutator of the normal-state Hamiltonian in the presence of external symmetry breaking fields \cite{RamiresPRB2016,RamiresPRB2018}. 
Remarkably, the effects of disorder on the SC state can also be inferred directly from the superconducting fitness function, if one introduces an impurity scattering potential to the normal-state Hamiltonian.

\begin{flushleft}
{\bf Robust superconductivity in the Bi$_2$Se$_3$-based materials}
\end{flushleft}

\begin{figure}[h]
    \includegraphics[width=11cm]{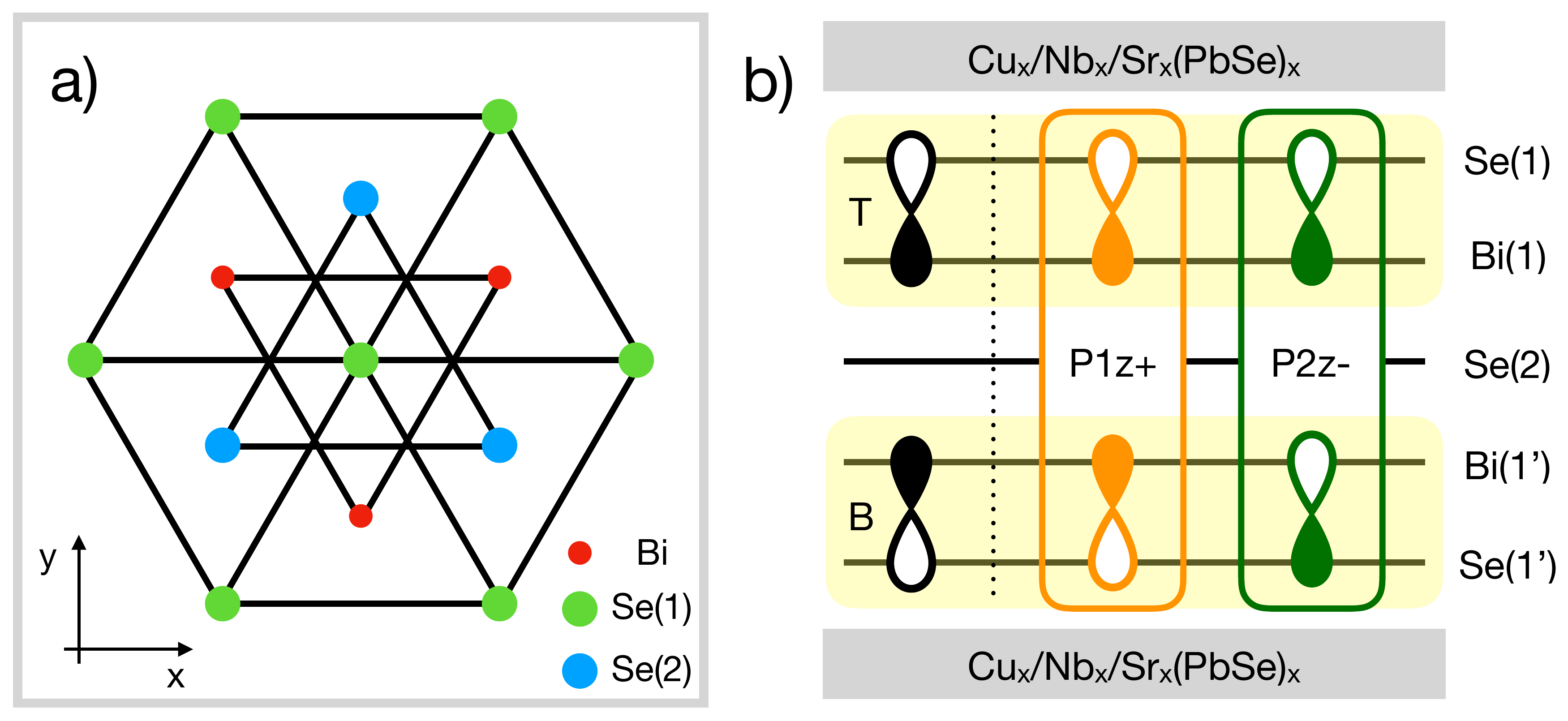}
    \caption{ \label{Fig:Structure} {\bf Material under consideration.} (a) Schematic representation of the crystal structure for materials in the family of Bi$_2$Se$_3$ (view along the $c$-axis); the gray rectangle depicts the reduced (monoclinic) symmetry in CPSBS. (b) Side view of the QL unit, highlighting the specific choice of orbitals: Shown on the left are the top (T) and bottom (B) layer orbitals used by Fu and Berg \cite{Fu2010}; shown on the right are the even (P1z$+$) and odd (P2z$-$) parity orbitals used in this work, identified as symmetric and antisymmetric superpositions of the orbitals in the top/bottom layers.}
\end{figure}

We can now use the fitness function to discuss the robustness of the SC state observed in the Bi$_2$Se$_3$-based materials. The normal state can be described by focusing on the quintuple-layer (QL) units, as schematically depicted in Fig. \ref{Fig:Structure}. The QL has $D_{3d}$ point group symmetry and the low-energy electronic structure can be described by an effective two-orbital model \footnote{ For the specific case of CPSBS, the (PbSe)$_5$ layers actually have square symmetry, such that the entire structure has the reduced point group symmetry $C_{2h}$. This is represented by the gray square in Fig. \ref{Fig:Structure} (a). More details in the Supplementary Materials}. The orbitals stem from Bi and Se atoms and have $p_z$ character. By a combination of hybridization, crystal field effects, and spin-orbit coupling (SOC), one can identify two effective orbitals with opposite parity, labeled $P_{1z+}$ and $P_{2z-}$, with the $\pm$ sign indicating the parity \cite{ZhangNP2009,LiuPRB2010}. A schematic representation of the orbitals is given in Fig. \ref{Fig:Structure}(b). In the basis $\Phi^\dagger_\bk  = (c_{1\uparrow}^\dagger, c_{1\downarrow}^\dagger, c_{2\uparrow}^\dagger, c_{2\downarrow}^\dagger)_\bk$, the normal-state Hamiltonian can be parametrized as Eq. \ref{Eq:H0}. In the presence of time-reversal and inversion symmetries, only the terms with $(a,b) = \{(0,0), (2,0),(3,0),(1,1),(1,2),(1,3)\}$ are allowed in the Hamiltonian. The properties of the respective matrices under the point group operations allow us to associate each of these terms to a given irreducible representation of $D_{3d}$, therefore constraining the momentum dependence of the form factors $h_{ab}(\bk)$. More details on the parametrization of the Hamiltonian are given in the Supplementary Materials.

The gap matrix can be parametrized as in Eq. \ref{Eq:Delta} in the orbital basis. As already noted, we focus on $\bk$-independent $\Delta$ matrices in this basis, because the pairing force is considered to be isotropic in Bi$_2$Se$_3$-based superconductors \cite{Fu2010}. Within the $D_{3d}$ point group symmetry, the allowed order parameters are summarized in Table S2 in the Supplementary Materials. In Fig. \ref{Fig:Pair}(a) we provide a schematic representation of pairing in the orbital basis, in which one can distinguish intra-orbital pairing in the even $A_{1g}$ channel from inter-orbital pairing in the odd channels \footnote{Note that the explicit form of the order parameters differ from the ones in Fu and Berg \cite{Fu2010} since the character of the orbitals is different in our formalism, which is evident from the form of the parity operator. Here we choose to start with a basis in which the parity operator is diagonal, which led to the insights described here. The two descriptions are in fact consistent, and are related by a unitary transformation \cite{LiuPRB2010}}. Given the experimental evidence for nodes along the $y$ direction in CPSBS \cite{Andersen2018}, here we focus on the following E$_u$ order parameter:
\begin{eqnarray}\label{Eq:DEu}
\Delta = \Delta_0 [i\tau_2 \otimes \sigma_1 (i\sigma_2)] = 
\Delta_0 \begin{pmatrix}
0 & 0 & -1 & 0 \\
0 & 0 & 0 & 1 \\
1 & 0 & 0 & 0 \\
0 & -1 & 0 & 0
\end{pmatrix},
\end{eqnarray}
which is spin triplet and orbital singlet. Under the action of the parity operator $P = \tau_3 \otimes \sigma_0$, one can infer that this is an odd-parity superconductor, even though the gap matrix is momentum independent. The oddness of this order parameter stems from the different parity of the two underlying orbitals. This $\bk$-independent order parameter in the orbital basis acquires nodes along the $y$ axis once projected to the Fermi surface in the band basis (Fig. \ref{Fig:Pair}(b), see Supplementary Materials for explicit calculations).

\begin{figure}[t]
    \includegraphics[width=14cm]{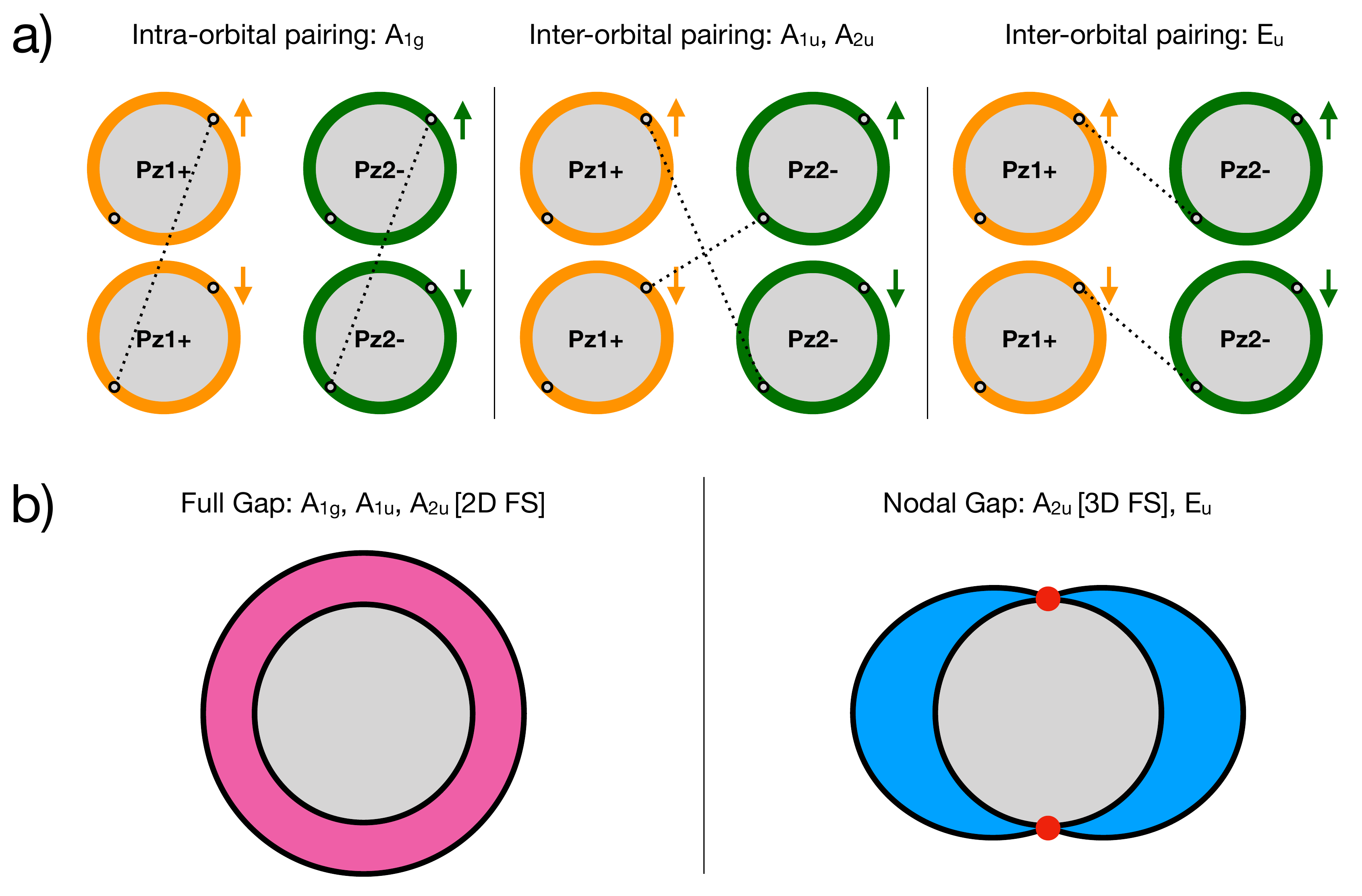}
    \caption{ \label{Fig:Pair} {\bf Possibilities of pairing.} (a) Schematic representation of the gap structure in the orbital basis. The yellow and green colors correspond to $P_{1z+}$ $P_{2z-}$ orbitals, respectively, as shown in Fig. \ref{Fig:Structure}. The dotted lines represent pairing between electrons with opposite momenta. Left: intra-orbital singlet pairing for A$_{1g}$. Middle: inter-orbital triplet/singlet pairing for A$_{1u}$/A$_{2u}$. Right: inter-orbital triplet pairing for $E_u$. (b) Schematic representation of the gap function in the band basis. Left: fully gapped, for order parameters in $A_{1g}$ and $A_{1u}$, as well as in $A_{2u}$ for a 2D Fermi surface (FS). Right: nodal gap structure for order parameters in $A_{2u}$ (for a 3D FS) and $E_u$. The red dot indicates the position of the nodes which can be read off from Table S3 in the Supplementary Materials. For a 3D FS these are point nodes on an ellipsoidal FS, while for a 2D FS these are line nodes extending along the $z$-direction on a cylindrical FS.}
\end{figure}

Given the order parameter shown above, we can now use the superconducting fitness function to understand the robustness of the SC state in CPSBS, for which we need to consider the explicit form of the matrix impurity scattering potential $V(\bk-\bk')$. The key aspect of the scattering potential for this material is the absence of orbital mixing, which is guaranteed by the opposite parity of the effective orbitals. This situation can be understood as follows: When we visualize the impurity as a local potential profile $v(\br)$ in real space, the magnitude of the scattering between two different orbitals is proportional to the overlap $ \int_\br \langle \phi_{P1z+} (\br) | v(\br) | \phi_{P2z-} (\br)\rangle$, which is zero within the assumption of a symmetric impurity potential. Hence, the scattering potential can only have the form
\begin{eqnarray}\label{Eq:V}
V (\bk-\bk') = \Phi^\dagger_\bk [ V_0 (\bk-\bk') \tau_0 \otimes \sigma_0 
+ V_s (\bk-\bk') \mathbf{S} \cdot (\tau_0 \otimes \boldsymbol{\sigma}) ]  \Phi_{\bk'},
\end{eqnarray}
where $V_a (\bk-\bk')$ is the Fourier transform of the potential scattering introduced by distinct sets of localized random impurities in real space. Here $a=\{0,s\}$ indicates nonmagnetic and magnetic impurity scattering, respectively; $\mathbf{S}$ signifies the spin of the magnetic impurities and $\boldsymbol{\sigma}$ the spin of the scattered electrons. 

Note that the scattering associated with nonmagnetic impurities is rather trivial in its matrix form, $\sim \tau_0 \otimes \sigma_0$, which always commute with the gap matrices $\Delta$ and leads to a zero fitness function, $F_C(\bk)=0$. As a consequence, the effective scattering rate for nonmagnetic impurities in CPSPB with the SC order parameter in the E$_u$ channel is zero, even though nodes are present in the excitation spectrum. Note that in this theoretical framework, the gap nodes are induced by the normal-state band structure once one translates the problem from the orbital basis to the band basis, as schematically shown in Fig. \ref{Fig:Pair}(b) and discussed in detail in the Supplementary Materials. In fact, the conclusion of zero scattering rate is valid for any SC order parameter possible for the Bi$_2$Se$_3$-based materials, because the identity matrix $\tau_0 \otimes \sigma_0$ commutes with any $\Delta$ of the form $\tau_a\otimes\sigma_b$.

Previous theoretical works have considered the effects of impurities in superconductors derived from Bi$_2$Se$_3$. Michaeli and Fu discussed how spin-orbit locking could parametrically protect unconventional SC states, but their results are valid only for states with pairs of electrons of the same chirality, restricting the analysis to order parameters in the $A_{1g}$ and $A_{1u}$ representations \cite{Michaeli2012}. More recently, Nagai proposed that the inter-orbital spin-triplet state with $E_u$ symmetry can be mapped to an intra-orbital spin-singlet $s$-wave pairing if the roles of spin and orbital are exchanged in the Hamiltonian, and he argued that this provides a mechanism for Anderson's theorem to remain valid when the spin-orbit coupling is strong \cite{Nagai2015}. Both works rely on assumptions which are not valid for all SC symmetry channels and depend on strong spin-orbit coupling. Such restrictions are not required for the above generalization of Anderson's theorem for multi-DOF superconductors, which shows that the robustness of the SC state against impurities is guaranteed by the isotropic nature of the pairing interaction written in the local orbital basis (leading to a momentum-independent order parameter in this microscopic basis), under the requirement that impurity scattering is not allowed between orbitals with opposite parity. These considerations are concisely captured by the superconducting fitness function $F_C(\bk)$. We emphasize that this framework has a particular importance in the context of topological superconductors, because the topological nature is often endowed by the extra DOF \cite{Sato2017}.

\begin{flushleft}
{\bf The Case of CPSBS}
\end{flushleft}

CPSBS is a superconductor obtained by intercalating Cu into its parent compound (PbSe)$_{5}$(Bi$_{2}$Se$_{3}$)$_{6}$, which is a member of the (PbSe)$_{5}$(Bi$_{2}$Se$_{3}$)$_{3m}$ homologous series realizing a natural heterostructure formed by a stack of the trivial insulator PbSe and the topological insulator Bi$_{2}$Se$_{3}$ \cite{segawa2015}. It was recently elucidated \cite{Andersen2018} that CPSBS belongs to the class of unconventional superconductors \cite{Fu2014, Ando2015, Sato2017} derived from Bi$_{2}$Se$_{3}$, including Cu$_x$Bi$_{2}$Se$_{3}$ \cite{Hor2010, Sasaki2011, Matano2016, Yonezawa2016}, Sr$_x$Bi$_2$Se$_3$ \cite{Liu2015, Nikitin2016, Pan2016, Du2017, Kuntsevich2018, Smylie2018}, and Nb$_x$Bi$_2$Se$_3$ \cite{Qiu2015, Shen2017, Asaba2017}, that possess a topological odd-parity SC state which spontaneously breaks rotation symmetry. Importantly, in contrast to the fully-opened gap in Cu$_x$Bi$_{2}$Se$_{3}$ \cite{Kriener2011}, the gap in CPSBS appears to have symmetry-protected nodes \cite{Andersen2018}. 

\begin{figure}[b]
	\centering
	\includegraphics[width=11cm]{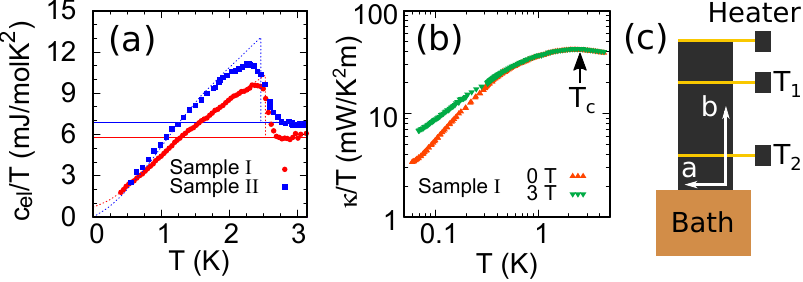}
	\caption{{\bf Specific heat and thermal conductivity across ${\bf T_c}$.} (a) Temperature dependencies of the electronic specific heat $c_{\rm el}$ of samples I and II (symbols), together with the theoretical curve for a line-nodal SC gap in the clean limit \cite{Won1994} assuming the SC volume fraction of 85\% and 100\%, respectively; horizontal lines correspond to $\gamma_{\rm el}$. Note that despite the strong scatterings in these samples, the clean-limit theory describes the $c_{\rm el}(T)$ data well, which is related to the robustness of the SC state against impurities. (b) Double-logarithmic plot of $\kappa/T$ vs $T$ for sample I measured in 0 and 3 T. (c) Schematics of the steady-state thermal-conductivity measurement setup.}
	\label{Fig:Exp1}
\end{figure}
 
\begin{figure}[b]
	\centering
	\includegraphics[width=10cm]{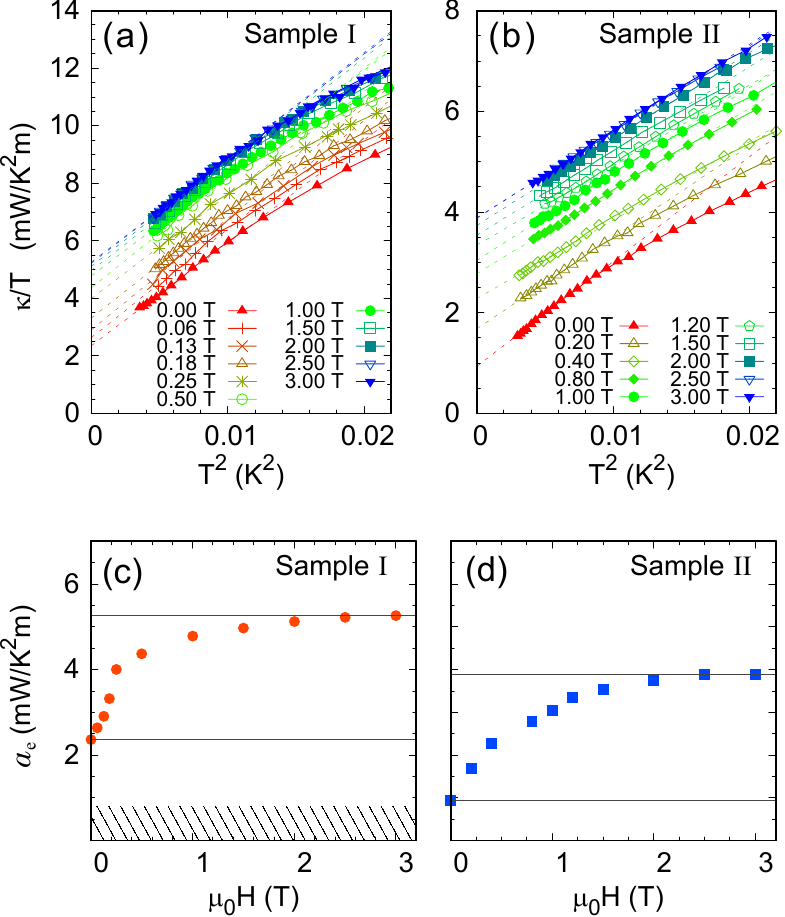}
	\caption{{\bf Ultra-low-temperature thermal conductivity.} (a,b) Plots of $\kappa/T$ vs $T^2$ for samples I and II measured in perpendicular magnetic fields up to 3 T. Dashed lines are the linear fits to the lowest-temperature part of the data; the intercept of these lines on the $\kappa/T$ axis gives $\kappa_0/T$.  (c,d) Magnetic-field dependencies of the electronic heat-transport coefficient $a_{\rm e}$ in samples I and II; solid lines mark the range of its change from 0 T to the normal state. The hatch at the bottom of panel (c) represents the expected background contributed by the non-SC portion of sample I.}
	\label{Fig:Exp2}
 \end{figure}

Figure \ref{Fig:Exp1}(a) shows the temperature dependence of the electronic specific heat $c_{\rm el}$, which is obtained from the total specific heat $c_p$ by subtracting the phononic contribution $c_{\rm ph}$ \cite{Andersen2018}, for the two samples studied in this work. The line-nodal gap theory \cite{Won1994} describes the $c_{\rm el}(T)$ data well, and the fits using this theory allow us to estimate the SC volume fraction, which is 85\% and 100\% for samples I and II, respectively. The thermal conductivity $\kappa$ was measured on the same samples down to 50 mK [Figs. \ref{Fig:Exp1}(b) and \ref{Fig:Exp2}] with the configuration depicted in Fig. \ref{Fig:Exp1}(c). Note that our previous study of $c_p$ in CPSBS in rotating magnetic field has revealed that line nodes are located in the $a$ direction \cite{Andersen2018}. 
The $c_p(T)$ data in the normal state obey $c_p = \gamma_{\rm el}T + \beta_{\rm ph} T^3$ \cite{Andersen2018} and we extract the phononic specific-heat coefficient $\beta_{\rm ph}$ = 5.1 (5.2) mJ/molK$^4$ and the electronic specific-heat coefficient $\gamma_{\rm el}$ = 5.8 (6.9) mJ/molK$^2$ for sample I (II). The $\kappa/T$ data present no anomaly at $T_c$ [Fig. \ref{Fig:Exp1}(b)], suggesting that electron-electron scattering is not dominant. 

In the $\kappa(T)$ data, one can separate the phononic and the electronic contributions to the heat transport when the $\kappa/T$ vs $T^2$ plot shows a linear behavior at low enough temperature. In our samples, this happens for $T \lesssim$ 100~mK [Figs. \ref{Fig:Exp2}(a) and \ref{Fig:Exp2}(b)], where phonons enter the boundary scattering regime and the phononic thermal conductivity $\kappa_{\rm ph}$ changes as $b_{\rm ph} T^3$. A finite intercept of the linear behavior in this plot means that there is residual electronic thermal conductivity $\kappa_0$ contributed by residual quasiparticles whose contribution increases linearly with $T$, i.e., $\kappa_0 = a_{\rm e}T$. In nodal superconductors, it has been established \cite{Hussey2002} that impurity scattering gives rise to a finite density of residual quasiparticles even at zero temperature, which is responsible for the finite $a_{\rm e}$. Upon application of a magnetic field $H$, vortices create additional quasiparticles which affects $\kappa$. In both samples, the magnetic-field dependence of $a_{\rm e}$ is sublinear, see Figs. \ref{Fig:Exp2}(c) and \ref{Fig:Exp2}(d), and this is most likely due to the Doppler shift of the superfluid around vortices, which leads to a $\sim\sqrt{H}$ increase in $c_{\rm el}$ in a nodal superconductor \cite{Volovik1993}. Note that the exact $H$ dependence of $a_{\rm e}$ would not be simple, because vortices enhance both the quasiparticle density {\it and} their scattering rate \cite{Ando2002, Vorontsov2006}. In 2.5 T, the superconductivity is fully suppressed and the $\kappa/T$ data are those of the normal state.

At this point, it is important to notice that these $\kappa/T$ data unambiguously show the presence of residual mobile quasiparticles down to 50 mK, which gives convincing evidence for the existence of gap nodes. In particular, sample II is essentially 100\% superconducting as indicated by the $c_p$ data, and yet, this sample in 0 T shows significant electronic heat conduction in the zero-temperature limit, which accounts for $\sim$24\% of the normal-state heat conduction [see Fig. \ref{Fig:Exp2}(d)]. This is impossible for a fully-gapped superconductor. The case for sample I is similar: Although the SC volume fraction of this sample is $\sim$85\% and hence one would expect some residual heat conduction at the level of 15\% of the normal-state value [shown by the hatch at the bottom of Fig. \ref{Fig:Exp2}(c)] due to the non-SC portion of the sample, the actual residual heat conduction in 0 T accounts for $\sim$45\% of the normal-state value, which strongly points to the contribution of residual nodal quasiparticles. 

To put the observed magnitude of $\kappa$ into context, the Wiedemann-Franz law $\kappa_0/T = L_0/\rho_{\rm res}$ is useful ($L_0 = \frac{\pi^2}{3} k_{\rm B}^2/e^2$  = 2.44 $\times$ 10$^{-8}$~$\Omega$W/K$^2$ is the Sommerfeld value of the Lorenz number and $\rho_{\rm res}$ is the residual resistivity). Using this formula and the observed $\kappa_0/T$ values in the normal state, we obtain $\rho_{\rm res}$ of 4.6 and 6.3 $\mu\Omega$m for samples I and II, respectively, which compares well to the direct measurements of $\rho_{\rm res}$ \cite{Sasaki2014}. 
We now make an order-of-magnitude estimate of the scattering time $\tau$ from $\rho_{\rm res}$ using the simple Drude model $\rho_{\rm res} = m^*/(ne^2\tau)$ and the relation between the effective mass $m^*$ and $\gamma_{el}$ for a two-dimensional free electron gas. With $\gamma_{\rm el}$ = 6.9 mJ/molK$^2$ of sample II, one obtains $m^* = (3 \hbar^2 \gamma_{\rm el} c_0)/(\pi V_{\rm mol}k_{\rm B}^2)$ = 4.7$m_{\rm e}$, where $V_{\rm mol}$ = 115.8 cm$^3$/mol is the Bi$_2$Se$_3$ molar volume used for the normalization of $c_p$ and $c_0$ = 1.27 nm is the height of the corresponding unit cell. With the typical carrier density 1.2 $\times$ 10$^{21}$ cm$^{-3}$ in CPSBS \cite{Sasaki2014}, one obtains $\tau$ = 2.2 $\times$ 10$^{-14}$ s for sample II. Since $m^*$ = 4.7$m_{\rm e}$ obtained from $\gamma_{el}$ is likely an overestimate of the transport effective mass, it only gives an upper bound for $\tau$. Hence, we obtain a lower bound of the scattering rate $\hbar \Gamma$ = $\hbar/\tau$ = 30 meV, which is already more than an order of magnitude larger than the SC gap $\Delta_0 \simeq$ 0.5 meV. We note that the mobility in CPSBS is only $\sim$10 cm$^2$/Vs, which precludes the determination of $m^*$ from quantum oscillations, although $m^* \simeq$ 0.2$m_{\rm e}$ has been estimated from quantum oscillations in Cu$_x$Bi$_2$Se$_3$ \cite{Lawson2012} and Nb$_x$Bi$_2$Se$_3$ \cite{Lawson2016}. Note that, if the actual effective mass is lighter than 4.7$m_{\rm e}$ in CPSBS, $\hbar \Gamma$ becomes larger and the conclusion about the robustness becomes even stronger. The estimates of $\hbar \Gamma$ for other Bi$_2$Se$_3$-based superconductors from the same Drude analyses unanimously give values larger than $\Delta_0$ (see Supplementary Materials), indicating the universal nature of the robustness in this family of unconventional superconductors.

It is crucial to notice that the universal thermal conductivity \cite{Lee1993, Graf1996, Taillefer1997, Proust2002}, which is expected only in clean superconductors satisfying $\hbar \Gamma \ll \Delta_0$, is {\it not} observed here.
A simple estimate of the expected magnitude of the universal thermal conductivity $\kappa_0^{\rm univ}$ given by $\kappa_{0}^{\rm univ}/T \approx (\gamma_{\rm el}v_{\rm F}^2 \hbar)/(2V_{\rm mol}\Delta_0)$ \cite{Graf1996} makes this situation clear: By using the Fermi velocity $v_{\rm F} = 4.8 \times 10^5$~m/s obtained from the angle-resolved photoemission experiments on CPSBS \cite{Nakayama2015}, one finds $\kappa_{0}^{\rm univ}/T \approx$ 8 W/K$^2$m, which is three orders of magnitude larger than the actual $\kappa_{0}/T$ in CPSBS in 0 T, indicating that the $\kappa_{0}/T$ value is significantly reduced from its clean-limit value due to strong impurity scattering. This gives convincing evidence that the strong scattering corresponding to $\hbar \Gamma \gg \Delta_0$ is at work not only in the normal state but also in the SC state. Note that in high-$T_c$ cuprates, the strong scattering leading to the ``bad metal'' behavior in the normal state is suppressed in the SC state, leading to the universal thermal conductivity to be observed in the mK region; clearly, this is not the case here. 

Hence, one can safely conclude that in CPSBS the energy scale of the scattering rate is much larger than the SC gap, which would normally preclude the realization of unconventional superconductivity with a nodal gap.  This provides a spectacular proof of the generalized Anderson's theorem in a multi-DOF superconductor. It is useful to note that the unusual robustness in $T_c$ against disorder was already noted for Cu$_x$Bi$_2$Se$_3$ \cite{Kriener2012} and Nb$_x$Bi$_2$Se$_3$ \cite{Smylie2017}, and the penetration-depth measurements of Nb$_x$Bi$_2$Se$_3$ also found evidence for nodes \cite{Smylie2016}, but the origin of the robustness remained a mystery. This mystery has actually been a reason for hindering some people from accepting Bi$_2$Se$_3$-based materials as well-established unconventional superconductors.  The present work finally solved this mystery, and it further provides a new paradigm for understanding the robustness of unconventional superconductivity. 
The new framework presented here will form the foundation for understanding the superconductivity in novel quantum materials where extra internal DOF such as orbitals, sublattices, or valleys govern the electronic properties.



\bibliography{Robustness_arxiv}

\vspace{5mm}

{\bf Acknowledgments} 
This work was funded by the Deutsche Forschungsgemeinschaft (DFG, German Research Foundation) - Project number 277146847 - CRC 1238 (Subprojects A04 and B01).

{\bf Additional information}
Correspondence and requests for materials should be addressed to A.R. (ramires@pks.mpg.de) and Y.A. (ando@ph2.uni-koeln.de).



\section*{Supplementary material}

## Materials and Methods

High-quality CPSBS single crystals were grown by a modified Bridgman method as described before [1]. Two samples from the same growth batch were measured. The dimensions of samples I and II were 2.8 × 2.5 × 0.35 and 5.6 × 2.0 × 0.20 mm$^3$, respectively. The exact $x$ values of samples I and II were 1.47 and 1.29, respectively. The specific heat $c_p$ was measured with a relaxation method in a Quantum Design PPMS down to 300 mK. Following previous works on CPSBS [1, 2], the SC volume fraction was estimated from the $c_p$ data by subtracting the phononic contribution $c_{\rm ph}$ and fitting the electronic contribution $c_{\rm el}$ with a line-nodal gap theory [3], yielding 85% and 100% for samples I and II, respectively. The shielding fraction at 1.8 K measured with a SQUID magnetometer in 0.2 mT applied parallel to the $ab$ plane was 75% and 88% in samples I and II, respectively. The thermal conductivity $\kappa$ was measured on the same samples in a dilution refrigerator (Oxford Instruments Kelvinox 400) with the standard steady-state method depicted in Fig. 3(c) in the main text using RuO$_2$ thermometers. The temperature gradient $\nabla T$ was applied parallel to the $b$ axis and the magnetic field was applied along the $c^*$ axis; note that CPSBS belongs to $C2/m$ space group, where $\vec{a} \perp \vec{b}$ and $\vec{c^*} \parallel \vec{a} \times \vec{b}$.

## The normal state Hamiltonian for materials in the family of Bi$_2$Se$_3$

The normal state Hamiltonian can be described by two effective orbitals of opposite parity, referred as $P_{1z+}$ and $P_{2z-}$ [4, 5]. In the basis $\Phi_{\mathbf{k}}^{\dagger} = (c_{1\uparrow}^{\dagger}, c_{1\downarrow}^{\dagger}, c_{2\uparrow}^{\dagger}, c_{2\downarrow}^{\dagger})_{\mathbf{k}}$, the Hamiltonian can be parametrized as:

$$H_0(\mathbf{k}) = \sum_{a,b} \Phi_{\mathbf{k}}^{\dagger} [h_{ab}(\mathbf{k}) \tau_a \otimes \sigma_b] \Phi_{\mathbf{k}}, \tag{1}$$

where $\tau_{a=1,2,3}$ are Pauli matrices encoding the orbital degrees of freedom (DOF), $\sigma_{b=1,2,3}$ are Pauli matrices encoding the spin DOF, and $\tau_0$ and $\sigma_0$ are two-dimensional identity matrices in orbital and spin space, respectively. In the presence of time-reversal and inversion symmetries, the only allowed terms in the Hamiltonian have the subscripts $(a, b) = \{(0,0), (2,0), (3,0), (1,1), (1,2), (1,3)\}$. Note that the time-reveral operator in our representation is $\Theta = K\tau_0 \otimes (i\sigma_2)$, where $K$ stands for complex conjugation, and the parity operator is $P = \tau_3 \otimes \sigma_0$. The properties of the $\tau_a \otimes \sigma_b$ matrices under the point group operations allow us to associate each of these terms to a given irreducible representation of $D_{3d}$, therefore

| $(a,b)$ | P | $C_3$ | $C_2'$ | $\sigma_d$ | Irrep | $h_{ab}(\mathbf{k})$ |
|---|---|---|---|---|---|---|
| $(0,0)$ | +1 | +1 | +1 | +1 | $A_{1g}$ | $C_0 + C_1 k_z^2 + C_2(k_x^2 + k_y^2)$ |
| $(2,0)$ | -1 | +1 | -1 | +1 | $A_{2u}$ | $B_0 k_z$ |
| $(3,0)$ | +1 | +1 | +1 | +1 | $A_{1g}$ | $M_0 + M_1 k_z^2 + M_2(k_x^2 + k_y^2)$ |
| $(1,1)$ | -1 | y | -1 | +1 | $E_u$ | $-A_0 k_y$ |
| $(1,2)$ | -1 | x | +1 | -1 | $E_u$ | $A_0 k_x$ |
| $(1,3)$ | -1 | +1 | +1 | -1 | $A_{1u}$ | - |

TABLE S1. Parametrization of the normal-state Hamiltonian given in Eq. 1 for materials in the family of Bi$_2$Se$_3$. The columns labeled by $P$, $C_3$, and $C_2'$ indicate how the basis matrices $\tau_a \otimes \sigma_b$, indicated by $(a,b)$, transform under the respective point group operations, such that one can associate these with different irreducible representations of $D_{3d}$ (Irrep). The last column gives the expansion of the form factors $h_{ab}(\mathbf{k})$ for small momentum.

constraining the momentum dependence of the form factors $h_{ab}(\mathbf{k})$ by symmetry. Here we take as generators the rotation along the $z$-axis by $2\pi/3$, $C_3 = \tau_0 \otimes e^{i\pi\sigma_3/3}$, and a rotation by $\pi$ along the $x$-axis, $C_2' = i\tau_3 \otimes \sigma_1$. It is also useful to define the $yz$-mirror transformation $\sigma_d = i\tau_0 \otimes \sigma_2$ in order to connect with the discussion for $C_{2h}$ symmetry below. Table S1 provides the details on the properties of each term in the normal state Hamiltonian and an expansion of $h_{ab}(\mathbf{k})$ for small momenta.

Note that the three first terms in Table S1 are spin-independent: $(0,0)$ and $(3,0)$ are even and associated with intra-orbital hopping, while $(2,0)$ is odd and encodes inter-orbital hopping. The last three terms are spin-dependent and inter-orbital in character, therefore all odd. The terms $(1,1)$ and $(1,2)$ are for a two-dimensional (2D) representation. The term $(1,3)$ is associated with trigonal warping of the Fermi surface and is usually dropped from the effective Hamiltonians since these carry at least terms of third order in momenta. For the chosen basis above, trigonal warping would only appear due to inter-layer hopping and would require terms of fifth order in momenta. For CPSBS, the coefficients $C_1$, $B_0$, and $M_1$ are smaller than the ones used for doped Bi$_2$Se$_3$ materials. As a consequence the Fermi surface for CPSBS is cylindrical (2D Fermi surface), while the Fermi surface for doped Bi$_2$Se$_3$ materials is usually ellipsoidal (3D Fermi surface).

| Irrep | Spin | Orbital | Parity | Matrix Form | Explicit Form |
|---|---|---|---|---|---|
| $A_{1g}$ | Singlet | Trivial | Even | $\tau_0 \otimes \sigma_0(i\sigma_2)$ | $c_{1\uparrow}c_{1\downarrow} + c_{2\uparrow}c_{2\downarrow}$ |
|  |  |  |  | $\tau_3 \otimes \sigma_0(i\sigma_2)$ | $c_{1\uparrow}c_{1\downarrow} - c_{2\uparrow}c_{2\downarrow}$ |
| $A_{1u}$ | Triplet | Singlet | Odd | $\tau_2 \otimes \sigma_3(i\sigma_2)$ | $c_{1\uparrow}c_{2\downarrow} + c_{1\downarrow}c_{2\uparrow}$ |
| $A_{2u}$ | Singlet | Triplet | Odd | $\tau_1 \otimes \sigma_0(i\sigma_2)$ | $c_{1\uparrow}c_{2\downarrow} - c_{1\downarrow}c_{2\uparrow}$ |
| $E_u$ | Triplet | Singlet | Odd | $i\tau_2 \otimes \sigma_1(i\sigma_2)$ | $c_{1\uparrow}c_{2\uparrow} - c_{1\downarrow}c_{2\downarrow}$ |
|  |  |  |  | $\tau_2 \otimes \sigma_2(i\sigma_2)$ | $c_{1\uparrow}c_{2\uparrow} + c_{1\downarrow}c_{2\downarrow}$ |

TABLE S2. Superconducting order parameters for materials in the family of $Bi_2Se_3$. Here we focus on the momentum-independent order parameters and highlight the associated irreducible representation (Irrep) and the spin, orbital character, and parity of the respective gap matrix. We write the order parameter in the matrix form $[\tau_a \otimes \sigma_b(i\sigma_2)]$ and show the explicit operator form. Here we factor out $(i\sigma_2)$ so that one can directly relate $b = 0$ to a singlet state and $b = \{1, 2, 3\}$ with the $\{x, y, z\}$ components of the d-vector parametrization for triplet states.

### The order parameters for materials in the family of $Bi_2Se_3$

The different gap matrices can be parametrized in a similar way. As motivated in the main text, here we already focus on isotropic, momentum-independent, gap matrices:

$$\Delta = \sum_{a,b} \Phi_{\mathbf{k}}^\dagger [d_{ab}\tau_a \otimes \sigma_b(i\sigma_2)]\Phi_{-\mathbf{k}}^*. \tag{2}$$

The allowed momentum-independent gap matrices can be determined by first searching for matrices satisfying $\Delta = -\Delta^T$, following from the fermionic nature of the electrons forming the Cooper pair. This condition implies that the matrices $[\tau_a \otimes \sigma_b(i\sigma_2)]$ associated with the order parameters need to be anti-symmetric. There are six such matrices. Furthermore, one can evaluate how these transform under the point-group symmetry operations defined above for the normal state, such that we can associate these with different irreducible representations of $D_{3d}$. The result of this analysis is displayed in Table S2.

These order parameters are constructed in the orbital basis, so one needs to transform these to the band basis in order to discuss the presence of nodes and their locations. Starting with the normal state Hamiltonian $H_0(\mathbf{k})$ in the orbital basis, one can always find a unitary

transformation $U(\mathbf{k})$ which diagonalizes the problem, taking it to the band basis:

$$H_B(\mathbf{k}) = U(\mathbf{k})H_0(\mathbf{k})U^\dagger(\mathbf{k}). \tag{3}$$

The unitary transformation can be explicitly written as

$$U(\mathbf{k}) = \begin{pmatrix} \frac{(h_{11}+ih_{12})(h_{30}-h_v)}{(h_{11}^2+h_{12}^2+h_{13}^2+h_{20}^2)N_+} & \frac{(h_{13}-ih_{20})(h_v-h_{30})}{(h_{11}^2+h_{12}^2+h_{13}^2+h_{20}^2)N_+} & 0 & \frac{1}{N_+} \\ \frac{(h_{13}+ih_{20})(h_{30}-h_v)}{(h_{11}^2+h_{12}^2+h_{13}^2+h_{20}^2)N_+} & \frac{(h_{11}-ih_{12})(h_{30}-h_v)}{(h_{11}^2+h_{12}^2+h_{13}^2+h_{20}^2)N_+} & \frac{1}{N_+} & 0 \\ \frac{(h_{11}+ih_{12})(h_{30}+h_v)}{(h_{11}^2+h_{12}^2+h_{13}^2+N_+^2)N_-} & -\frac{(h_{20}-ih_{20})(h_{30}+h_{13})}{(h_{11}^2+h_{12}^2+h_{13}^2+h_{20}^2)N_-} & 0 & \frac{1}{N_-} \\ \frac{(h_{13}+ih_{20})(h_{30}+h_v)}{(h_{11}^2+h_{12}^2+h_{13}^2+h_{20}^2)N_-} & \frac{(h_{11}-ih_{30})(h_{30}+h_v)}{(h_{12}^2+h_{12}^2+h_{13}^2+h_{20}^2)N_-} & \frac{1}{N_-} & 0 \end{pmatrix}, \tag{4}$$

where $h_v = |\mathbf{h}|$, with $\mathbf{h} = (h_{20}, h_{30}, h_{11}, h_{12}, h_{13})$ a five-dimensional vector and

$$N_\pm = \sqrt{\frac{2}{1 \pm \frac{h_{30}}{h_v}}}. \tag{5}$$

Here we suppressed the momentum dependences in the parameters $h_{ab}$ for clarity.

This transformation gives $H_B(\mathbf{k}) = \text{diag}[E_0^+(\mathbf{k}), E_0^+(\mathbf{k}), E_0^-(\mathbf{k}), E_0^-(\mathbf{k})]$ with eigenvalues

$$E_0^\pm(\mathbf{k}) = h_{00}(\mathbf{k}) \pm |\mathbf{h}(\mathbf{k})|. \tag{6}$$

Note that the states are doubly degenerate, given the presence of time-reversal and inversion symmetries.

If we are interested in the superconducting properties in the band basis, we can use the same unitary transformation to rotate the order parameter matrix

$$\Delta_B(\mathbf{k}) = U(\mathbf{k})\Delta[U^\dagger(-\mathbf{k})]^*, \tag{7}$$

from which we can infer the structure of the gap. For the specific case of the $E_u$ order parameter realized in CPSBS, which has the following form in the orbital basis

$$\Delta = \Delta_0[i\tau_2 \otimes \sigma_1(i\sigma_2)] = \Delta_0 \begin{pmatrix} 0 & 0 & -1 & 0 \\ 0 & 0 & 0 & 1 \\ 1 & 0 & 0 & 0 \\ 0 & -1 & 0 & 0 \end{pmatrix}, \tag{8}$$

we can rotate it to the band basis to find

$$\Delta_B(\mathbf{k}) = \frac{\Delta_0}{h_v}\begin{pmatrix} -ih_{20} & ih_{12} & \frac{ih_{20}h_{30}}{h_p} & \frac{h_{11}h_v - ih_{12}h_{30}}{h_p} \\ ih_{12} & ih_{20} & -\frac{h_{11}h_v + ih_{12}h_{30}}{h_p} & -\frac{ih_{20}h_{30}}{h_p} \\ \frac{ih_{20}h_{30}}{h_p} & -\frac{h_{11}h_v + ih_{12}h_{30}}{h_p} & ih_{20} & -ih_{12} \\ \frac{h_{11}h_v - ih_{12}h_{30}}{h_p} & -\frac{ih_{20}h_{30}}{h_p} & -ih_{12} & -ih_{20} \end{pmatrix}, \tag{9}$$

5

where $h_p = \sqrt{h_{11}^2 + h_{12}^2 + h_{13}^2 + h_{20}^2}$. Here, again, we suppressed the $\mathbf{k}$-dependences of the $h_{ab}$ functions for clarity, but they are in fact $\mathbf{k}$-dependent as shown in Table S1.

Note that the gap matrix in the band basis has now both intra-band components (diagonal blocks) and inter-band components (off-diagonal blocks), so one cannot generally think of a simple gap structure at the Fermi level. In this band basis, it is interesting to note that the gap matrix has contributions from both the upper and lower parts of the gapped Dirac cone (note the relative sign of the diagonal blocks), and the gap magnitude for this $E_u$ order parameter depends on $h_{12}(\mathbf{k}) \sim A_0 k_x$ and $h_{20}(\mathbf{k}) \sim B_0 k_z$. These form factors determine the presence and position of the nodes, which in this cases lies on the $y$-axis.

At this point we believe it is more enlightening to look directly at the eigenvalues of the Bogoliubov-de Gennes Hamiltonian:

$$H_{BdG}(\mathbf{k}) = \begin{pmatrix} H_0(\mathbf{k}) & \Delta \\ \Delta^\dagger & -H_0^*(-\mathbf{k}) \end{pmatrix}, \tag{10}$$

where $H_0(\mathbf{k})$ is the normal-state Hamiltonian and $\Delta = \Delta_0[\tau_c \otimes \sigma_d(i\sigma_2)]$, with different irreducible representations associated with different subscripts $(c, d)$, as listed in Table S2. Here $\Delta_0$ is the magnitude of the order parameter. The eigenvalues are explicitly written as

$$E^{\pm,\pm}(\mathbf{k}) = \pm\sqrt{h_{00}^2(\mathbf{k}) + \mathbf{h}^2(\mathbf{k}) + \Delta_0^2 \pm 2\sqrt{h_{00}^2(\mathbf{k})\mathbf{h}^2(\mathbf{k}) + \Delta_0^2 f_C^2(\mathbf{k})}}, \tag{11}$$

where the function

$$f_C^2(\mathbf{k}) = \frac{1}{16}\text{Tr}[\tilde{F}_C^{0\dagger}(\mathbf{k})\tilde{F}_C^0(\mathbf{k})], \tag{12}$$

is written in terms of the normalized superconducting fitness function $\tilde{F}_C^0(\mathbf{k}) = F_C^0(\mathbf{k})/\Delta_0$:

$$F_C^0(\mathbf{k}) = H_0(\mathbf{k})\Delta - \Delta H_0^*(-\mathbf{k}), \tag{13}$$

which is different for each order parameter presented in Table S2.

Note that the normal-state Fermi surface satisfies $E_0^\pm(\mathbf{k}) = 0$, or better $h_{00}^2(\mathbf{k}) = \mathbf{h}^2(\mathbf{k})$, such that the energy spectra in the superconducting state at the momenta belonging to the normal-state Fermi surface can be written as [6–8]

$$E^{\pm,\pm}(\mathbf{k})\Big|_{FS} = \pm\sqrt{2h_{00}^2(\mathbf{k}) + \Delta_0^2 \pm 2\sqrt{h_{00}^4(\mathbf{k}) + \Delta_0^2 f_C^2(\mathbf{k})}}. \tag{14}$$

If we now expand for small $\Delta_0$, we find

$$E^{\pm,\pm}(\mathbf{k})\Big|_{FS} \approx \pm\sqrt{2h_{00}^2(\mathbf{k}) + \Delta_0^2 \pm 2h_{00}^2(\mathbf{k}) \pm \frac{\Delta_0^2}{h_{00}^2(\mathbf{k})}f_C^2(\mathbf{k})}, \tag{15}$$

such that

$$E^{\pm,+}(\mathbf{k})\Big|_{FS} \approx \pm\sqrt{4h_{00}^2(\mathbf{k}) + \Delta_0^2\left(1 + \frac{f_C^2(\mathbf{k})}{h_{00}^2(\mathbf{k})}\right)} \qquad (16)$$

and

$$E^{\pm,-}(\mathbf{k})\Big|_{FS} \approx \pm\Delta_0\sqrt{\left(1 - \frac{f_C^2(\mathbf{k})}{h_{00}^2(\mathbf{k})}\right)}. \qquad (17)$$

The last equation corresponds to the low lying states in the superconducting phase, such that we can define the gap structure in the band basis as the function:

$$\mathcal{D}_B(\mathbf{k}) = E^{+,-}(\mathbf{k})\Big|_{FS} - E^{-,-}(\mathbf{k})\Big|_{FS} \approx 2\Delta_0\sqrt{\left(1 - \frac{f_C^2(\mathbf{k})}{h_{00}^2(\mathbf{k})}\right)}. \qquad (18)$$

This definition is in direct analogy to the single band problem in which the dispersion in the normal state is $\epsilon(\mathbf{k})$, and in the superconducting state is $E^{\pm} = \pm\sqrt{\epsilon^2(\mathbf{k}) + \Delta^2}$. Hence we can identify the gap by taking $\epsilon(\mathbf{k}) = 0$, or $E^{\pm}|_{FS} = \pm|\Delta|$. This form indicates that there can be nodes in the spectra of CPSBS given the nontrivial form factors originating from the normal-state Hamiltonian $H_0(\mathbf{k})$, even though the gap in the orbital basis is isotropic. Note that the fitness function $F_C^0(\mathbf{k})$ appears again in this context, proving its importance for the understanding of the phenomenology of superconducting states in complex materials.

From the equation above, it is clear that the condition for the presence of nodes is:

$$h_{00}^2(\mathbf{k}) - f_C^2(\mathbf{k}) = 0. \qquad (19)$$

We evaluate this quantity for all order parameters listed in Table S2 in order to infer the presence of nodes and their positions for each case. The results are summarized in Table S3 below. Note that the position of the nodes is determined purely by the parameters $h_{ab}(\mathbf{k})$ given in the normal-state Hamiltonian.

**Analysis for $C_{2h}$ symmetry**

For the specific case of CPSBS, the $(PbSe)_5$ layers actually have square symmetry, such that the entire structure has the reduced point group symmetry $C_{2h}$. This is represented by the gray rectangle in Fig. 1(a) in the main text. Now $C_3$ is not a symmetry transformation anymore, but $C_2'$ and $\sigma_d$ are still valid symmetry operations which allow us to generate the character table for the reduced group. Given the presence of parity and time-reversal

7

| Irrep | Matrix Form | $h_{00}^2(\mathbf{k}) - f_C^2(\mathbf{k})$ | Explicit form | Zero at 3D FS | Zero at 2D FS |
|---|---|---|---|---|---|
| $A_{1g}$ | $\tau_0 \otimes \sigma_0(i\sigma_2)$ | $h_{00}^2(\mathbf{k})$ | $[C_0 + C_1 k_z^2 + C_2(k_x^2 + k_y^2)]^2$ | - | - |
| | $\tau_3 \otimes \sigma_0(i\sigma_2)$ | $h_{30}^2(\mathbf{k})$ | $[M_0 + M_1 k_z^2 + M_2(k_x^2 + k_y^2)]^2$ | - | - |
| $A_{1u}$ | $\tau_2 \otimes \sigma_3(i\sigma_2)$ | $h_{20}^2(\mathbf{k}) + h_{11}^2(\mathbf{k}) + h_{12}^2(\mathbf{k})$ | $B_0^2 k_z^2 + A_0^2(k_x^2 + k_y^2)$ | - | - |
| $A_{2u}$ | $\tau_1 \otimes \sigma_0(i\sigma_2)$ | $h_{11}^2(\mathbf{k}) + h_{12}^2(\mathbf{k})$ | $A_0^2(k_x^2 + k_y^2)$ | $k_x = k_y = 0$ | - |
| $E_u$ | $i\tau_2 \otimes \sigma_1(i\sigma_2)$ | $h_{20}^2(\mathbf{k}) + h_{12}^2(\mathbf{k})$ | $B_0^2 k_z^2 + A_0^2 k_x^2$ | $k_z = k_x = 0$ | $k_x = 0$ |
| | $\tau_2 \otimes \sigma_2(i\sigma_2)$ | $h_{20}^2(\mathbf{k}) + h_{11}^2(\mathbf{k})$ | $B_0^2 k_z^2 + A_0^2 k_y^2$ | $k_z = k_y = 0$ | $k_y = 0$ |

TABLE S3. Analysis of the gap structure for materials in the family of Bi$_2$Se$_3$. The parameter determining the nodes, $h_{00}^2(\mathbf{k}) - f_C^2(\mathbf{k})$, is written in terms of $h_{ab}(\mathbf{k})$ and explicitly for small momenta. The 3D Fermi surface (FS) should be understood as an ellipsoidal Fermi surface around the Γ point, while a 2D FS should be understood as a cylindrical FS along the $z$-axis.

symmetries, the only terms allowed in $H_0(\mathbf{k})$ are the same as the ones enumerated for the case of $D_{3d}$ symmetry, but now these are mapped to different irreducible representations as follows: $A_{1g/u} \to A_{g/u}$, $A_{2g/u} \to B_{g/u}$, $E_{g/u} \to \{A_{g/u}, B_{g/u}\}$. The last correspondence means that the two dimensional irreducible representations of $D_{3d}$ are split in $C_{2h}$. The consequence for the normal state is that the parameter $A_0$ does not need to be the same for the $(1, 1)$ and $(1, 2)$ terms (see Table S1). The same correspondences are valid for the order parameters listed in Table S2. Note that there are only one-dimensional representations in $C_{2h}$, and therefore the notion of nematic states, strictly speaking, does not apply.

**The concept of superconducting fitness and the effective scattering rate**

For the derivation of Eq. 4 in the main text, we follow the standard treatment for the determination of the critical temperature for superconductors in the presence of impurities [9, 10]. We start with the Bogoliubov-de Gennes Hamiltonian:

$$H_{BdG}(\mathbf{k}) = \Psi_{\mathbf{k}}^{\dagger} \begin{pmatrix} H_0(\mathbf{k}) & \Delta(\mathbf{k}) \\ \Delta^{\dagger}(\mathbf{k}) & -H_0^*(-\mathbf{k}) \end{pmatrix} \Psi_{\mathbf{k}}, \qquad (20)$$

written in terms of a multi-DOF Nambu spinor $\Psi_{\mathbf{k}}^{\dagger} = (\Phi_{\mathbf{k}}^{\dagger}, \Phi_{-\mathbf{k}}^T)$, encoding several degrees of freedom (DOF) within $\Phi_{\mathbf{k}}^{\dagger} = (c_{1\mathbf{k}\uparrow}^{\dagger}, c_{1\mathbf{k}\downarrow}^{\dagger}, ..., c_{n\mathbf{k}\uparrow}^{\dagger}, c_{n\mathbf{k}\downarrow}^{\dagger})$, as defined in the main text. Here $H_0(\mathbf{k})$ is the normal-state Hamiltonian and $\Delta(\mathbf{k})$ the superconducting order parameter.

From this matrix Hamiltonian one can define the Green's function (GF) for the clean system as

$$G_w^0(\mathbf{k}) = (i\omega I_{2n} - H_{BdG}(\mathbf{k}))^{-1}, \tag{21}$$

where $I_{2n}$ is the 2n-dimensional identity matrix and $n$ is the number of DOF, which are doubled in the Nambu basis.

The effects of impurities can be accounted for by the introduction of a scattering potential. For the specific case of CPSBS, we can write the explicit form of the scattering potential in the Nambu basis as

$$V_{BdG}(\mathbf{k} - \mathbf{k}') = V_0(\mathbf{k} - \mathbf{k}')\rho_3 \otimes \tau_0 \otimes \sigma_0 + V_s(\mathbf{k} - \mathbf{k}')\mathbf{S} \cdot (\tau_0 \otimes \boldsymbol{\alpha}), \tag{22}$$

where

$$\boldsymbol{\alpha} = \left(\frac{1+\rho_3}{2}\right) \otimes \boldsymbol{\sigma} + \left(\frac{1-\rho_3}{2}\right) \otimes \sigma_2\boldsymbol{\sigma}\sigma_2, \tag{23}$$

with the Pauli matrices $\rho_i$ encoding particle and hole spaces. $V_a(\mathbf{k}-\mathbf{k}')$ is the Fourier transform of the potential scattering introduced by distinct sets of localized random impurities in real space. Here $a = \{0, s\}$ indicates nonmagnetic and magnetic impurity scattering, respectively, $\mathbf{S}$ corresponds to the spin of the magnetic impurities and $\boldsymbol{\sigma}$ to the spin of the scatterred electrons.

Introducing a given realization of impurities in principle removes the translational invariance present in the clean system, but after averaging over random impurity locations the GF recovers translational invariance and one can then propose as ansatz the following renormalized GF:

$$G_w(\mathbf{k}) = \left(i\tilde{\omega} I_{2n} - \tilde{H}_{BdG}(\mathbf{k})\right)^{-1}, \tag{24}$$

where inside $\tilde{H}_{BdG}(\mathbf{k})$ only the gap matrix is renormalized to $\tilde{\Delta}(\mathbf{k})$. The two renormalized parameters, $\tilde{\omega}$ and $\tilde{\Delta}(\mathbf{k})$, are determined self-consistently by the introduction of a self energy $\Sigma_w(\mathbf{k})$ associated with the effects of impurity scattering:

$$[G_w(\mathbf{k})]^{-1} = \left[G_w^0(\mathbf{k})\right]^{-1} - \Sigma_w(\mathbf{k}). \tag{25}$$

The self energy is calculated within the Born approximation as

$$\Sigma_w(\mathbf{k}) = n_i \int_{\mathbf{k}'} V_{BdG}(\mathbf{k} - \mathbf{k}')G_w(\mathbf{k}')V_{BdG}(\mathbf{k} - \mathbf{k}'), \tag{26}$$

9

where $n_i$ is the impurity concentration.

Within this framework, the renormalized parameters $\tilde{\omega}$ and $\tilde{\Delta}_0$ (the magnitude of the gap, from here on assumed to be momentum independent) follow the coupled equations:

$$\tilde{\omega} = \omega + \frac{1}{2}\left(\frac{1}{\tau_1} + \frac{1}{\tau_2}\right)\frac{\tilde{\omega}}{\sqrt{\tilde{\omega}^2 + \tilde{\Delta}_0^2}}, \tag{27}$$

$$\tilde{\Delta}_0 = \Delta_0 + \frac{1}{2}\left(\frac{1}{\tau_1} + \frac{1}{\tau_2}\right)\frac{\tilde{\Delta}_0}{\sqrt{\tilde{\omega}^2 + \tilde{\Delta}_0^2}} - \pi N(0)\frac{\tilde{\Delta}_0}{\sqrt{\tilde{\omega}^2 + \tilde{\Delta}_0^2}}\frac{1}{4}\left\langle\overline{Tr[\tilde{F}_C^\dagger(\Omega_{\mathbf{k}})\tilde{F}_C(\Omega_{\mathbf{k}})]}\right\rangle_{\mathbf{k}}, \tag{28}$$

in which we can identify the normalized fitness function $\tilde{F}_C(\mathbf{k}) = F_C(\mathbf{k})/\Delta_0$, defined as

$$F_C(\mathbf{k} - \mathbf{k}') = V(\mathbf{k} - \mathbf{k}')\Delta - \Delta V^*(\mathbf{k} - \mathbf{k}'), \tag{29}$$

with $V(\mathbf{k} - \mathbf{k}')$ acting only in the particle space, as defined in Eq. 8 in the main text.

For the equations above we further define the standard scattering rates:

$$\frac{1}{2\tau_1} = \frac{1}{2}n_i\pi N(0)\int d\cos\theta' |V_0(\theta')|^2, \tag{30}$$

$$\frac{1}{2\tau_2} = \frac{1}{2}n_i\pi N(0)S(S+1)\int d\cos\theta' |V_s(\theta')|^2,$$

where $N(0)$ is the density of states at the Fermi energy in the normal state, and $S$ is the magnitude of the spin of the impurities.

The effects of impurities on the critical temperature can be evaluated from the linearized gap equation. For that, the ratio $u = \tilde{\omega}/\tilde{\Delta}_0$ is a useful parameter. From Eqs. 27 and 28 above, one can write the ratio $\omega/\Delta_0$ in terms of $u$ as

$$\frac{\omega}{\Delta_0} = u\left(1 - \frac{\hbar\Gamma_{\text{Eff}}}{\Delta_0}\frac{1}{\sqrt{1+u^2}}\right), \tag{31}$$

where we already identified the effective scattering rate:

$$\hbar\Gamma_{\text{Eff}} = \frac{1}{4}\left\langle\overline{Tr[\tilde{F}_C^\dagger(\Omega_{\mathbf{k}})\tilde{F}_C(\Omega_{\mathbf{k}})]}\right\rangle_{\mathbf{k}}. \tag{32}$$

Note that if $\hbar\Gamma_{\text{Eff}} = 0$ we find $\frac{\tilde{\omega}}{\tilde{\Delta}_0} = \frac{\omega}{\Delta_0}$, such that there is no renormalization of the parameters defining the critical temperature and therefore $T_c$ is not suppressed.

Let us now evaluate the the critical temperature explicitly. In the absence of impurities, we can determine the critical temperature $T_c^0$ from the self-consistent equation

$$\Delta_{\mathbf{k}} = -vk_BT\sum_n\int_{\mathbf{k}}\frac{1}{4}Tr[\rho_1\sigma_2\hat{G}_{\omega_n}^0(\mathbf{k})], \tag{33}$$

where $v$ is the pairing strength, $k_B$ the Boltzmann constant, $T$ the temperature, and $n$ an index for the Matsubara frequencies $\omega_n = (2n+1)\pi k_B T$. Writing the GF explicitly, one finds the familiar form for a momentum-independent gap function

$$1 = v k_B T \sum_n \int_{\mathbf{k}} \frac{1}{\omega_n^2 + \xi_{\mathbf{k}}^2 + \Delta_0^2}, \tag{34}$$

which leads to the equation defining the critical temperature:

$$\frac{1}{N(0)v} = \log\left(\frac{4e^\gamma}{\pi} \frac{\omega_D}{2 k_B T_c^0}\right), \tag{35}$$

after taking the sum over momenta to an integral over energy by the introduction of a density of states $N(\xi)$. Here $\gamma$ is the Euler's constant and $\omega_D$ is the energy cutoff around the FS within which there is pairing.

The gap equation in the presence of impurities become

$$\Delta_{\mathbf{k}} = -v k_B T \sum_n \int_{\mathbf{k}} \frac{1}{4} Tr[\rho_1 \sigma_2 \hat{G}_{\omega_n}(\mathbf{k})], \tag{36}$$

or

$$\Delta_0 = v k_B T \sum_n \int_{\mathbf{k}} \frac{\tilde{\Delta}_0}{\tilde{\omega}_n^2 + \xi_{\mathbf{k}}^2 + \tilde{\Delta}_0^2}, \tag{37}$$

with renormalized parameters in the right hand side. Taking a constant DOS near the FS, one may rewrite it to

$$\Delta_0 = N(0) v k_B T \sum_n \int_{-\omega_D}^{\omega_D} d\xi \frac{\tilde{\Delta}_0}{\tilde{\omega}_n^2 + \tilde{\Delta}_0^2} \frac{1}{1 + \xi^2/(\tilde{\omega}_n^2 + \tilde{\Delta}_0^2)}. \tag{38}$$

The relabeling $x = \xi/\sqrt{\tilde{\omega}_n^2 + \tilde{\Delta}_0^2}$ leads to

$$\Delta_0 = N(0) v k_B T \sum_n \frac{\tilde{\Delta}_0}{\sqrt{\tilde{\omega}_n^2 + \tilde{\Delta}_0^2}} \int_{-x_D}^{x_D} dx \frac{1}{1 + x^2} \tag{39}$$

$$= N(0) v k_B T \sum_n \frac{\tilde{\Delta}_0}{\sqrt{\tilde{\omega}_n^2 + \tilde{\Delta}_0^2}} 2 ArcTan\left(\frac{\omega_D}{\sqrt{\tilde{\omega}_n^2 + \tilde{\Delta}_0^2}}\right)$$

$$\approx N(0) v \pi k_B T \sum_n \frac{\tilde{\Delta}_0}{\sqrt{\tilde{\omega}_n^2 + \tilde{\Delta}_0^2}},$$

where in the last step we used that $ArcTan(...) \to \pi/2$ for large arguments.

At this point it is illuminating to rewrite this equation in terms of the ratio $u_n = \tilde{\omega}_n/\tilde{\Delta}_0$:

$$\Delta_0 = N(0) v \pi k_B T \sum_n \frac{1}{\sqrt{1 + u_n^2}}. \tag{40}$$

11

From the self-consistent equations above, we already identified the ratio $\omega_n/\Delta_0$ in Eq. 31. Close to $T_c$, with $\Delta_0 \to 0$, we can rewrite this ratio in the Matsubara frequency notation as

$$\frac{\omega_n}{\Delta_0} \approx u_n - \frac{\hbar \Gamma_{\text{Eff}}}{\Delta_0} \tag{41}$$

or

$$\Delta_0 u_n \approx \omega_n + \hbar \Gamma_{\text{Eff}}, \tag{42}$$

and the gap equation becomes

$$\Delta_0 = N(0)v\pi k_B T \sum_n \frac{1}{|u_n|} \tag{43}$$

or

$$1 = N(0)v\pi k_B T \sum_n \frac{1}{\omega_n + \hbar \Gamma_{\text{Eff}}}. \tag{44}$$

Writing the Matsubara frequencies explicitly, one obtains

$$\frac{1}{N(0)v} = \sum_{n>0} \frac{1}{(n+1/2) + \rho_c}, \tag{45}$$

where $\rho_c = \hbar \Gamma_{\text{Eff}}/2\pi k_B T_c$, defining the new $T_c$.

Given the identity of the pure system from Eq. 35, we can write

$$\sum_{n>0} \frac{1}{(n+1/2)} - \log\left(\frac{4e^\gamma}{\pi} \frac{\omega_D}{2k_B T_c}\right) = 0. \tag{46}$$

By adding it to Eq. 45 above, one obtains

$$\frac{1}{N(0)v} = \sum_{n>0} \left\{ \frac{1}{(n+1/2) + \rho_c} - \frac{1}{(n+1/2)} \right\} + \log\left(\frac{4e^\gamma}{\pi} \frac{\omega_D}{2k_B T_c}\right) \tag{47}$$

$$= \sum_{n>0} \left\{ \frac{1}{(n+1/2) + \rho_c} - \frac{1}{(n+1)} + \frac{1}{(n+1)} - \frac{1}{(n+1/2)} \right\} + \log\left(\frac{4e^\gamma}{\pi} \frac{\omega_D}{2k_B T_c}\right),$$

where we can identify the digamma function:

$$\Psi(z) = -\gamma + \sum_{n>0} \left\{ \frac{1}{(n+1)} - \frac{1}{(n+z)} \right\}. \tag{48}$$

We can finally obtain

$$\log\left(\frac{4e^\gamma}{\pi} \frac{\omega_D}{2k_B T_c^0}\right) = \Psi(1/2) - \Psi(1/2 + \rho_c) + \log\left(\frac{4e^\gamma}{\pi} \frac{\omega_D}{2k_B T_c}\right) \tag{49}$$

12

or

$$\log\left(\frac{T_c}{T_c^0}\right) = \Psi(1/2) - \Psi(1/2 + \rho_c), \tag{50}$$

where, again, $\rho_c = \hbar\Gamma_{\text{Eff}}/2\pi k_B T_c$ and

$$\hbar\Gamma_{\text{Eff}} = \frac{1}{4}\left\langle\overline{Tr[\tilde{F}_C^\dagger(\Omega_{\mathbf{k}})\tilde{F}_C(\Omega_{\mathbf{k}})]}\right\rangle_{\mathbf{k}} \tag{51}$$

is the effective scattering rate, written as a universal form in terms of the superconducting fitness function $F_C(\mathbf{k})$.

We would like to highlight that, even for arbitrarily large $\tau_1$ and $\tau_2$ in the normal state, if $F_C(\mathbf{k}) = 0$ the effective scattering rate $\hbar\Gamma_{\text{Eff}} = 0$ such that there is no renormalization of the ratio $\omega/\Delta_0$ and therefore no suppression of the critical temperature.

**Drude analysis of the scattering rates in Bi$_2$Se$_3$-based superconductors**

The estimates of the scattering rates are performed for Cu$_x$Bi$_2$Se$_3$, Sr$_x$Bi$_2$Se$_3$, and Nb$_x$Bi$_2$Se$_3$ in the same manner as was done for CPSBS by taking materials parameters from the literature. Table S4 summarized the estimates. If one puts the $\hbar\Gamma$ values in this

| Material | CPSBS | Cu$_{0.3}$Bi$_2$Se$_3$ | Sr$_{0.1/0.06}$Bi$_2$Se$_3$ | Nb$_{0.25}$Bi$_2$Se$_3$ |
|---|---|---|---|---|
| $\gamma$ (mJ/molK$^2$) | 6.90 | 1.79 [11] | 0.42 [12] | 4.54 [13] |
| $n$ (m$^{-3}$) | 1.2×10$^{27}$ | 1.2×10$^{26}$ [11] | 2.6×10$^{25}$ [14] | 1.5×10$^{26}$ [13] |
| $\rho_N$ ($\mu\Omega$m) | 6.3 | 2.02 [11] | 0.93 [14] | 1.70 [13] |
| $V_{\text{mol}}$ (cm$^3$/mol) | 115.8 | 85 [11] | ~85 | ~85 |
| Dimension | 2D | 3D | 2D [15] | 2D? |
| $k_F$ (nm$^{-1}$) | – | 1.57 | – | – |
| $m^*$ ($m_e$) | 4.7 | 2.60 [11] | 0.29 | 3.1 |
| $\tau$ (s) | 2.2×10$^{-14}$ | 3.5×10$^{-13}$ | 4.3×10$^{-13}$ | 4.4×10$^{-13}$ |
| $\hbar\Gamma$ ($=\hbar/\tau$) (meV) | 30 | 1.9 | 1.6 | 1.5 |
| $\Delta_0$ (meV) | ~0.5 | ~0.5 | ~0.5 | ~0.5 |
| $\hbar\Gamma/\Delta_0$ | ~60 | ~3.8 | ~3.1 | ~3.0 |

TABLE S4. Estimate of the scattering rates in Bi$_2$Se$_3$-based superconductors from the simple Drude analysis as was done for CPSBS in the main text.

table into $\hbar\Gamma_{\text{Eff}}$ in Eq. 50 (which is the same as Eq. 4 of the main text), one sees that the $T_c$ of CPSBS under this amount of scattering would be only 1% of the $T_c$ of a clean material, if it would have been a "conventional" nodal superconductor. For the other materials, the $T_c$ would be about 25% of the clean-case values.

### Quantized thermal conductivity due to surface Majorana fermions

Note that, while some recent theoretical works on the heat transport in topological superconductors focus on quantized thermal conductivity due to surface Majorana fermions [16, 17], the particle-hole symmetry in the SC state [18] dictates that the chemical potential is always pinned to the charge neutrality point of the Majorana cone, which in turn means that the Majorana fermions can contribute only a *single transport mode* to the heat transport. This single-mode nature is the source of the quantization, but it also means that the contribution from the surface Majorana fermions to the total thermal conductivity, $\kappa_0^{\text{Majorana}}$, is extremely small in a bulk sample (i.e. $\kappa_0^{\text{Majorana}}/T \sim 10^{-8}$ W/K$^2$m for a 0.1-mm-thick sample). In the case of a nodal topological superconductor [18], the thermal conductivity would be dominated by the bulk quasiparticles even in the zero-temperature limit, and thus one would not expect to see the contribution from surface Majorana fermions.

### Phononic contribution to the thermal conductivity

For completeness, we discuss the phononic contribution in $\kappa$. As shown in Fig. S1(a), the coefficient $b_{\text{ph}}$ in both samples are essentially magnetic-field independent. The different

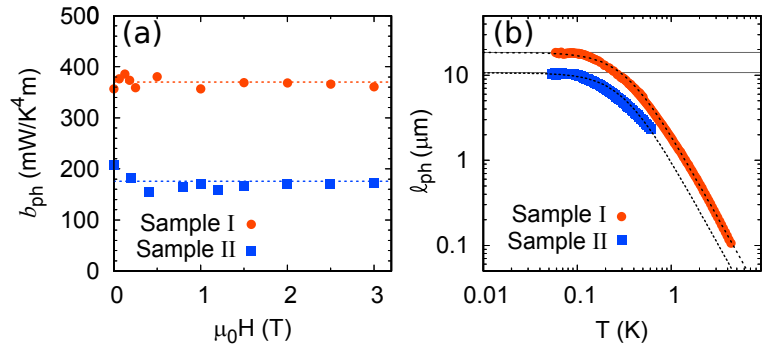

FIG. S1. **Behavior of phonons.** (a) The phononic thermal-conductivity coefficient $b_{\text{ph}}$ in samples I and II, which is essentially independent of $H$. (b) Temperature dependencies of the phonon mean free path $\ell_{\text{ph}}$ in samples I and II; horizontal lines mark the lowest-temperature saturated value, while dashed lines are the fits of $\ell_{\text{ph}}(T) = \ell_0 + c_2 T^{-2}$ to the data.

14

$b_{\rm ph}$ values signify a difference in the phonon mean free path $\ell_{\rm ph}$ in the boundary scattering regime. The $\kappa_{\rm ph}$ in this regime is described by $\kappa_{\rm ph} = \frac{1}{3}\beta_{\rm ph}T^3 v_s \ell_{\rm ph}/V_{\rm mol}$, where $v_s$ is the sound velocity. Using $\kappa_{\rm ph}/T^3 = b_{\rm ph} \approx 370$ (180) mW/K$^4$m for sample I (II), $\beta_{\rm ph} \approx 5$ mJ/molK$^4$ from the $c_p$ data, and $v_s \approx 1400$ m/s from the Debye model, we estimate $\ell_{\rm ph} \approx 19$ (11) $\mu$m. These values are much shorter than the effective sample size approximated by $w^* = (4wt/\pi)^{1/2} = 1.1$ (0.72) mm ($w$ and $t$ are the width and the thickness of the sample). This suggests that there are some internal defect structures which limit the phonon mean free path in our CPSBS samples. At the temperature range above the boundary scattering regime, the $T$-dependence of $\kappa_{\rm ph}$ allows one to infer the dominant scatterers. Our analysis shows that the leading $T$-dependent term in $\ell_{\rm ph}$ is the $T^{-2}$ term [Fig. S1(b)]. Such a term is primarily caused by scattering off stacking faults, which may well exist in a homologous-series material such as CPSBS.

---



\end{document}